# Selective addressing of solid-state spins at the nanoscale via magnetic resonance frequency encoding

**Authors:** H. Zhang[1,2]*, K. Arai[1,3]*, C. Belthangady[1,2], J.-C. Jaskula[1-3] and R. L. Walsworth[1,2,4],‡

**Affiliations:**

1. Harvard-Smithsonian Center for Astrophysics, Cambridge, Massachusetts 02138, USA.
2. Department of Physics, Harvard University, Cambridge, Massachusetts 02138, USA.
3. Department of Physics, Massachusetts Institute of Technology, Cambridge, Massachusetts 02139, USA.
4. Center for Brain Science, Harvard University, Cambridge, Massachusetts 02138, USA.

* These authors contributed equally to this work.

‡ Corresponding author (rwalsworth@cfa.harvard.edu)

**Abstract**

The nitrogen-vacancy (NV) centre in diamond is a leading platform for nanoscale sensing and imaging, as well as quantum information processing in the solid state. To date, individual control of two NV electronic spins at the nanoscale has been demonstrated. However, a key challenge is to scale up such control to arrays of NV spins. Here we apply nanoscale magnetic resonance frequency encoding to realize site-selective addressing and coherent control of a four-site array of NV spins. Sites in the array are separated by 100 nm, with each site containing multiple NVs separated by ~15 nm. Microcoils fabricated on the diamond chip provide electrically tuneable magnetic-field gradients ~0.1 G/nm. Tailored application of gradient fields and resonant microwaves allow site-selective NV spin manipulation and sensing applications, including Rabi oscillations, imaging, and nuclear magnetic resonance (NMR) spectroscopy with nanoscale resolution. Microcoil-based magnetic resonance of solid-state spins provides a practical platform for quantum-assisted sensing, quantum information processing, and the study of nanoscale spin networks.

**Introduction**

In recent years, nitrogen vacancy (NV) colour centres in diamond have been successfully applied to a wide range of problems in quantum information, sensing, and metrology in both the physical and life sciences[1]. For example, single NV centres have been used for a loophole-free Bell test of quantum realism[2], probing nanoscale phenomena in condensed matter systems[3-7], and NMR spectroscopy and imaging of nanoscale ensembles of nuclear spins[8-10] including single proteins[11] and individual proton spins[12]. Large ensembles of NV centres have provided magnetic imaging with combined micron-scale resolution and millimetre field-of-view, e.g., for mapping paleomagnetism in primitive meteorites[13] and ancient Earth rocks[14], genetic studies of magnetotactic bacteria[15,16], and identifying biomarkers in tumour cells[17]. However, it remains a challenge to realize the intermediate regime of mesoscopic arrays of NV spins with selective nanoscale addressing and coherent control of NVs at each site in the array. Such a capability could be a platform technology for applications such as high-spatial-dynamic-range magnetic imaging[18-20] and quantum–assisted sensing[21-23], as well as scalable quantum information processing[24,25] and simulation[26,27].

In the present work, we experimentally demonstrate selective coherent manipulation of an array of four NV spin sites, equally spaced by ~100 nm, using a frequency encoding technique inspired by magnetic resonance imaging (MRI). In frequency encoding, the positions of spins at different locations in a sample are mapped onto their resonance frequency using tuneable magnetic-field gradients; and then frequency-tailored pulse sequences address and control spins at specific target positions. This technique is widely employed in conventional biomedical MRI for image slice-selection with millimetre-scale resolution[28], and was recently used in trapped-ion experiments[29] to control coherently ions separated by a few microns. Creating strong and spatially homogeneous magnetic-field gradients that can be switched rapidly compared to the spin coherence lifetime is the primary technical challenge for multi-site spin control. For our NV-diamond experiment, we achieve a tuneable gradient strength ~0.1 G/nm over a $1.2 \times 8$ μm$^2$ area by use of a micrometre-scale electromagnetic coil (microcoil) fabricated onto the diamond by e-beam lithography (Fig. 1). The field gradient is spatially uniform to 5%, can be modulated at ~1 MHz, and its application requires no active cooling of the sample. Gradients of comparable magnitudes over such a large area and with comparable switching rates are difficult to achieve using ferromagnets (Supplementary Figs. 1, 2)[30]. We apply these electrically tuneable magnetic-field gradients to a series of demonstrations on the array of NV spins, including site-selective electron spin resonance (ESR) spectroscopy, Rabi oscillations, Fourier imaging[20], and nuclear magnetic resonance (NMR) spectroscopy, all with spatial resolution ≈30 nm.

**Results**

**Frequency-encoding of NV spin sites in diamond**

The frequency-encoding system for site-selective addressing of NV spins in diamond consists of fabricated arrays of NV centres and micron-scale coils, integrated into a home-built scanning confocal microscope as illustrated in Fig. 1a. NV centres are located at a depth of ≈20 nm from the surface of a [100]-cut diamond chip ($4 \times 4 \times 0.5$ mm$^3$) and are arranged in a two-dimensional array (Supplementary Fig. 3). A uniform, static magnetic field of $B_0 = 128$ G, created by an external Helmholtz coil pair, is applied along the diamond [111] direction, which corresponds to one of the four NV crystallographic orientations (Fig. 1a inset). A microwave antenna for coherent spin-state control and a magnetic-field gradient microcoil are fabricated directly on the diamond surface. A strong field gradient, aligned nominally with $B_0$, is created by sending electric currents through the microcoil in an anti-Helmholtz configuration. In each optically-addressable confocal volume, four NV sites spaced by 100 nm are created by mask implantation (Fig. 1b)[31]. Figure 1c shows the energy level diagram of the negatively charged NV centre. The ground-state spin triplet is optically initialized into spin state $|0\rangle$, coherently addressed with microwaves, and read out by spin-state-dependent fluorescence. On applying a magnetic field gradient, the NV spins in different sites acquire a position-dependent Zeeman splitting between the $|\pm 1\rangle$ states with well-resolved resonance frequencies. NV spins in each site can thus be selectively manipulated by microwaves with distinct, site-specific frequencies.

As shown by scanning electron microscope (SEM) image in Fig. 1d, the spacing between the gradient microcoil wires is 2.5 μm. In each NV site, there are approximately three NV centres with the same crystallographic orientation, as determined by the rate of fluorescence counts in a typical confocal volume (Fig. 1e), which is consistent with the estimated NV concentration of $5 \times 10^{11}$ cm$^{-2}$. See Methods and Supplementary Figs. 4, 5 for more details of the mask implantation and microcoil fabrication. Superresolution optical imaging with a stimulated-emission depletion (STED) microscope confirms the formation of four NV sites with a spacing of about 100 nm within a confocal volume (Fig. 1e, inset). More details of the STED imaging apparatus is found in the Methods section.

**Demonstration of site-selective NV spin addressing and control**

As a benchmark demonstration of site-selective NV addressing via frequency encoding, we performed optically-detected electron spin resonance (ESR) measurements with a DC electric current of 250 mA sent through the microcoil (Fig. 2a). Four ESR peaks, corresponding to the four NV sites in the array, are clearly observed (Fig. 2b). We fit the data to a sum of four Lorentzian curves, and determined the splitting between adjacent resonances to be $\Delta f = 29(3)$ MHz. From the SEM and STED images we find the mean separation between NV sites to be $\Delta x = 96(7)$ nm, and thus the field gradient to be $dB/dx = 2\pi \times \Delta f/\gamma\Delta x = 0.11(1)$ G nm$^{-1}$. Note that the observed ~30% variation in ESR peak linewidth is consistent with a simple model of inhomogeneous line broadening in the presence of the applied magnetic field gradient due to multiple (typically three) NV centres being randomly distributed in position within each site.

Next we demonstrated site-selective coherent Rabi driving of NV spins via the pulse sequence illustrated in Fig. 2c. After initializing all NV spins into the |0⟩ state with a 5 µs long green laser pulse, frequency encoding is instantiated with a DC field gradient, and a microwave pulse tuned to the ESR frequency of a target NV site is applied for a duration $\tau_{MW}$. Finally, another 5 µs laser pulse is applied to read out the NV spin states via a fluorescence measurement. NVs in the target site exhibit Rabi oscillations as the duration of the microwave pulse is varied. As the microwave frequency is adjusted to match the ESR frequency at each of the four sites, we obtain the data shown in Fig. 2d. The fidelity of such site-selective control of the NV spins via frequency encoding is estimated to be >97.4% (see Methods). By fitting each Rabi oscillation data set with a sinusoid, we determine the Rabi frequencies at all four NV sites, with results that are consistent with a numerical simulation of the microwave field produced by the antenna (Fig. 2e, see also Supplementary Discussion 1).

**Demonstration of site-selective NV imaging and NMR spectroscopy**

To illustrate the utility of the frequency encoding technique, we performed site-selective 1D imaging of the array of NV centres. The experimental protocol (Fig. 3a) integrates our previously-demonstrated NV Fourier imaging technique[20] with frequency encoding to resolve the array's sub-diffraction-limit spatial structure with site-selection capability. In analogy with conventional MRI, an alternating (AC) magnetic field gradient, synchronized with a Hahn echo NV pulse sequence, phase-encodes spatial information about the NV sites in wavenumber or '$k$-space' onto the NV spins' phase, while also isolating the NV spins from local magnetic field variations that induce dephasing. In particular, the array of NV spins with real-space positions $x_i$ ($i$ = A,…, D) is exposed to an AC gradient of magnitude $(dB/dx)_j$ and thus acquires a position-dependent phase $\varphi = 2\pi k_j x_i$, where $k_j = (2\pi)^{-1}\gamma\tau(dB/dx)_j$ defines the $j^{th}$ point in Fourier or $k$-space. Here, $\gamma/2\pi = 2.8$ MHz/G is the NV gyromagnetic ratio and $\tau$ is the total NV spin precession time in the Hahn echo sequence. The optically-detected NV signal for a point in $k$-space is proportional to the sum across all NV sites of the cosine of the acquired NV spin phase at each site: $s(k_j) \sim \sum_i \cos(2\pi k_j x_i)$. By incrementally stepping through a range of field gradient amplitudes with $\tau$ fixed, one measures the NV signal as a function of $k$ to yield a $k$-space image. In the following discussion we drop subscript $j$ for simplicity. The real-space image is then-reconstructed by a Fourier transformation of the $k$-space image: $S(x) = F[s(k)]$, where $abs[S(x)]$ gives the relative positions of the NV sites in the array. Note that the resolution of the real-space image is $(k_{max})^{-1}$, where $k_{max} = (2\pi)^{-1}\gamma\tau|dB/dx|_{max}$ is the maximum $k$ value used in the measurement. In the presence of an additional DC frequency-encoding field gradient and frequency-tuned microwave pulses, only NV centres at a specific target site in the array (e.g., $x_A$) are subject to the phase-encoding protocol and hence contribute to both the Fourier image (e.g., $s(k_j, x_A) \sim \cos(2\pi k_j x_A)$) and the real space image (e.g., $S(x_A)$).

We demonstrated this nanoscale NV imaging protocol using a phase-encoding magnetic field gradient of sinusoidal form $(dB/dx)_j = G_j \sin(2\pi t/\tau)$, where $G_j$ is the gradient magnitude for the $j^{th}$ point in $k$-space. Example 1D $k$-space and real-space NV images are shown in Figs. 3b and 3c, respectively, for 512 equally-spaced $k$-space points and the NV spin free precession time fixed at $\tau = 0.9$ μs. The maximum $k$-space value is $k_{max} = 0.021(1)$ nm$^{-1}$, which is induced by an AC gradient magnitude of $G_{max} = 0.0068(3)$ G nm$^{-1}$ corresponding to a microcoil current of $I = 25$ mA. $k_{max}$ implies a 1D real-space resolution of $\delta x = (2k_{max})^{-1} = 24(2)$ nm, which is much less than the 100 nm separation between sites in the array of NV spins, but is insufficient to resolve individual NVs within one site. Note that site-selective Hahn-echo measurements using the DC field gradient, but without phase-encoding, confirm that the coherence times ($T_2$) for NV centres in all array sites are ≥2 µs and thus decoherence is insignificant during the imaging protocol (see Figs. 4a and 4b and Supplementary Fig. 6).

The top images in both Figs. 3b and 3c (labeled ABCD) have no DC frequency-encoding gradient applied, and hence contain contributions from all four NV sites; whereas the bottom four images in these figures (labeled A, B, C or D) are acquired with site-selective frequency encoding using a DC magnetic field gradient of 0.1 G nm$^{-1}$. The four resolved signal peaks in image ABCD in Fig. 3c indicate the real-space locations of the NV sites in the array, with mean site separation = 93(1) nm and mean site diameter = 32(2) nm. These results are consistent with the NV array geometry determined using site-selective frequency-encoding (images A, B, C, and D in Fig. 3c), which collectively yield mean site separation = 92(1) nm and mean site diameter = 33(2) nm. Consistent results are also found for the $k$-space images with and without site-selective frequency-encoding. Specifically, site-selected $k$-space images A, B, C, and D in Fig. 3b exhibit coherent single-spatial-frequency oscillations, with the frequency at each site (determined from fits to the image data) varying linearly with location in the array, as expected for a uniform DC gradient field (Supplementary Fig. 7). $k$-space image ABCD, acquired without site-selection, displays spatial frequency beats in accord with a coherent sum of the $k$-space images A, B, C, D from the four NV sites.

As a final demonstration, we combined the frequency encoding technique with an AC magnetometry experimental protocol (Fig. 4a) to perform site-selective NMR spectroscopy of the $^{15}$N nuclear spins that are constituents of the NV centres at each site. With no applied DC field gradient, all four NV sites contribute to the measured Hahn echo signal, allowing determination of the NV ensemble $T_2 \geq 2$ µs (Fig. 4b, upper panel and Supplementary Fig. 6). With the frequency encoding DC gradient applied and the microwave frequency tuned to address only one NV site at a time, the Hahn echo signal is modulated by the $^{15}$N NMR signal at a Larmor precession frequency given by the transverse component of the gradient field $B_\perp$ at the selected site (Fig. 4b, lower panel and Supplementary Discussion 2). The measured $^{15}$N Larmor frequency as a function of the varying transverse field $B_\perp$ across multiple sites yields a $^{15}$N nuclear gyromagnetic ratio of $|\gamma_n/2\pi| = 0.46(4)$ kHz G$^{-1}$, which is consistent with the accepted value[19].

**Discussion**

Our application of magnetic resonance frequency-encoding to NV spins in diamond provides the first method for selective addressing and coherent control of a mesoscale array of solid-state spins with nanoscale resolution. The present technique, by which pulsed magnetic field gradients across the NV array are induced via electrically-tuneable microcoils fabricated on the diamond chip, should be extendable to multiple spatial dimensions, larger and denser NV arrays, and smaller length scales (Supplementary Discussion 3). In particular, an improvement in gradient strength to >1 G nm$^{-1}$ should be practical by increasing the electric current through the microcoil, optimally matching the microcoil's impedance to the current supply, and introducing an active temperature control system to better remove microcoil-induced heat from the diamond (Supplementary Discussion 4). This enhanced gradient field could allow selective addressing with >95% fidelity of a micron-scale array of dipolar-coupled NV centres, each spaced by ≈10 nm (Supplementary Discussions 4 and 5). Such a network of strongly-interacting spins with high spatial dynamic range has many potential applications, including in quantum sensing and imaging[32], quantum information processing[33,34], studies of quantum spin transport[35], and as quantum simulators for exotic quantum and topological phases (e.g., spin liquids and supersolids[36], quantum spin Hall effect[37], and topological insulators[38]). We also emphasize the simplicity and flexibility of the gradient microcoil design, which facilitates integration with other systems such as microfluidics and micro-electro-mechanical systems (MEMS). Furthermore, the frequency-encoding technique should be integratable with a wide range of NV sensing protocols, including for DC and AC magnetic fields, electric fields, and temperature. These advantages open new directions for applications, including wide-field NMR imaging of nanoscale nuclear spin diffusion and effusion in cellular or microfluidic environments via q-space detection[39], and spatially-selective nanoscale imaging of strongly correlated spins in condensed matter systems[36].

## Methods

**Creation of 2D NV centre arrays with mask implantation**

The diamond sample used in this experiment is an electronic-grade, single-crystal, and 99.999% $^{12}$C high-purity chemical vapour deposition (CVD) [100]-cut chip ($4 \times 4 \times 0.5$ mm$^3$) obtained from Element 6 Corporation. Registration markers are fabricated on the diamond substrate by e-beam lithography and reactive ion etching. All subsequent fabrication steps use the same spatial coordinates defined by these markers. A polymethyl methacrylate (PMMA) ion implantation mask is used to spatially control NV centre formation in a three-level hierarchical structure of 2D NV arrays (Supplementary Fig. 3). $^{15}$N$^+$ ions are implanted with a dose of $1 \times 10^{13}$ cm$^{-2}$ at an implantation energy of 14 keV. The conversion efficiency from nitrogen ions to NV centres after high-temperature vacuum annealing (1200 °C, 4 hours) is approximately 6%, which is determined by comparing the measured NV fluorescence signal from a confocal spot with that from a single NV centre. From simulations using the stopping and range of ions in matter (SRIM) program [40], the NV centres are estimated to be 21(7) nm below the diamond surface. Typical NV spin coherence times are $T_2^* \approx 580$ ns and $T_2 \approx 4.5$ μs.

**Fabrication of gradient microcoil and microwave antenna**

The magnetic field gradient microcoil and microwave antenna are fabricated on the diamond chip near the NV arrays (Supplementary Fig. 4). A double-layer PMMA process is employed to form an undercut, which improves the quality of copper lift-off. Here the bottom layer is PMMA 495k C9 and the top layer is 950k C4. A thin layer of Chromium (~10 nm) is deposited onto the PMMA stack to improve the surface conductance and reduce charging during e-beam lithography. An Elionix F-125 e-beam writing system is used for patterning with exposure dosage and beam energy set at 2,600 μC cm$^{-2}$ and 125 kV, respectively. A 30 nm Ti layer and a 970 nm Au layer are then deposited in an e-beam evaporator, followed by lift-off in MicroChem Remover PG solution. Electrical resistance measurements, performed using a four-probe station, give a resistance of ~2 Ω for each microcoil. A heat sink is then attached to the back surface of the diamond coverslip to enhance heat dissipation. (Supplementary Fig. 8)

**Confocal scanning laser microscopy**

Site-selective NV spin addressing and sensing experiments are performed using a custom-built confocal scanning laser microscope. A 400 mW diode-pumped solid state laser (Changchun New Industries) operating at 532 nm is used for NV optical excitation. An acousto-optic modulator (Isomet Corporation) modulated at 80 MHz is used for pulsing the laser beam. Laser pulses are sent to a 100x, 1.3 NA oil-immersion objective lens (Nikon CFI Plan Fluor). A three-axis motorized stage (Micos GmbH) allows scanning of the NV-containing diamond sample in the focal plane of the objective. Red fluorescence from NV centres is separated from the 532 nm excitation light with a dichroic beam-splitter (Semrock LM01-552-25), focused onto a single-mode fibre with a mode-field-diameter ~5 μm, and then collected by a silicon avalanche photodiode detector (Perkin Elmer SPCM-AQRH-12). Pulse signals from the detector is transmitted to a computer through a DAQ card (National Instrument PCI 6221).

**STED microscopy**

Superresolution NV optical images (e.g., Fig. 1e) are recorded on a homebuilt CW-STED microscope [41] Stimulated Emission Depletion (STED) microscopy is based on applying a strong, spatially structured optical depletion field to switch off peripheral fluorescent emitters through stimulated emission. In our system, the depletion field is a doughnut-shaped optical beam at 750 nm, which is applied to NV centres in the field of view at the same time as a 532 nm Gaussian-shaped excitation laser beam. The doughnut-shaped depletion beam rapidly drives to the ground electronic state all NV centres that are off the dark spot in the centre of the beam axis, while limiting unwanted NV re-excitation. Thus, only NVs on the beam axis produce NV fluorescence, which is imaged as the microscope is scanned. The doughnut beam is created by a vortex phase plate (RPC Photonics), which imprints a staircase $180^\circ$ phase shift on the input 750 nm laser beam such that diametrically opposite rays are out of phase. This results in creating a dark spot in the centre of the depletion beam by destructive interference at the focal point of the objective (100x TIRF, Nikon). Aberrations in the interference pattern are minimized via an optimization procedure in which the doughnut beam is imaged by reflection from 80 nm diameter gold nanoparticles; and the position of the vortex phase plate is determined, with respect to the 750 nm laser beam, that minimizes the intensity at the doughnut centre[42]. The STED point spread function is then imaged with ~1 µm-deep isolated single NV centres in the bulk diamond sample. An improvement in spatial resolution from about 250 nm (confocal) to 50 nm (STED) is routinely achieved by applying a 300 mW depletion doughnut beam while the excitation beam power is kept as low as 100 µW. With the same experimental parameters, a four site array of NV centres can be STED imaged across a 600 nm × 250 nm field-of-view with $100 \times 100$ pixels, at a speed of 10 ms/pixel, realizing a per pixel NV contrast-to-noise ratio of ~5. With such a STED imaging procedure, four individual NV sites, each separated by 96(7) nm are distinguished. 2D Gaussian filtering (4 × 1.5 pixels) and thresholding are applied to improve the image rendering, as in Fig. 1e.

**Characterization of the magnetic field gradient**

To characterize the performance of the microcoil, the magnetic field gradient is measured as a function of electric current through the microcoil by optically detected NV ESR. An external Helmhotz coil pair is used to apply a uniform magnetic field of $B_0$ =128 along the [111] diamond crystallographic orientation (on-axis), which leaves the NV spins along the other three orientations (off-axis) degenerate. A gradient magnetic field $B(x)$ is then introduced by sending an electric current $I$ through the microcoil. This gradient field is aligned nominally with $B_0$ and the [111] diamond crystallographic orientation, although there is also a modest transverse component to the gradient field $B_\perp(x)$ (Supplementary Discussion 2). The ground state Hamiltonian of the on-axis NV centres is thus expressed as $H/\hbar = DS_z^2 + \gamma B_0 S_z + \gamma B \cdot S$, where $\hbar$ is the reduced Planck constant, $\gamma = 2\pi \times 2.8$ MHz G$^{-1}$ is the NV gyromagnetic ratio and $S$ is the spin-1 operator. The change of NV fluorescence is recorded under continuous optical and microwave excitations. Supplementary Fig. 9a displays the results of the measurement, showing the change of NV fluorescence as a function of microwave carrier frequency $f$ and current $I$. Of the four observed resonance bands, the outer two correspond to the ESR transitions of on-axis NV centres while the inner two are the ESR transitions of three degenerate off-axis NV centres. The measurements are consistent with a simulation based on the above NV spin Hamiltonian (Supplementary Fig. 9b). As the current is increased, the resonance bands become broader because the four NV sites split. A higher resolution ESR scan of the highlighted region in Supplementary Fig. 9a quantifies the splitting of the resonance band for on-axis NV centres (Supplementary Fig. 9c). At around $I = 200$ mA, the band begins to clearly split into four peaks, corresponding to four proximal NV sites, and the ESR contrast decreases to ~2%. Consistency with simulation is again found (Supplementary Fig. 9d). By fitting the ESR spectrum with a curve comprised of four Lorentzian lineshapes for a given current value, the resonance frequency of each NV site is extracted $f_i$ ($i = 1,…,4$). Since the separation between NV sites is known to be $\Delta x =$ -96(7) nm from the SEM and STED images, the field gradient at each current value can be obtained using $dB/dx = 2\pi \times \Delta f/\gamma \Delta x$, where $\Delta f$ is the measured frequency splitting between adjacent resonance peaks. Repeating this analysis for all current values, the field gradient per unit current is found to be $dB/dx/I = 0.45(2)$ G nm$^{-1}$ A$^{-1}$ (Supplementary Fig. 10). In particular, at $I = 250$ mA, the measured field gradient is $dB/dx = 0.11(1)$ G nm$^{-1}$. A numerical simulation of the field gradient is performed in three steps. First, the magnetic field spatial distribution produced by the microcoil at a fixed current is simulated using COMSOL. Next, the magnetic field spatial distribution for the entire current range is determined under the assumption that the field is linearly proportional to the current. Finally, the ESR resonance peaks for all NV centres are calculated by diagonalising the ground state Hamiltonian with the obtained field distribution as an input. The resulting analysis yields a field gradient per unit current of $dB/dx/I = 0.48$ G nm$^{-1}$ A$^{-1}$, which is in reasonable agreement with the measured value (Supplementary Fig. 10).

**Estimation of Rabi driving fidelity**

From the site-selective Rabi driving measurements presented in Fig. 2, the fidelity of coherent NV spin driving in the presence of a DC gradient and with a Rabi frequency of $\Omega$ can be estimated by evaluating the off-resonant excitation ("crosstalk") error: $p_{\text{err}} \approx p_{\text{dip}} + p_{T_1} + p_{T_2} + p_{\text{off}}$ [43]. The first term, $p_{\text{dip}} \sim (\kappa/\Omega)^2 \sim 10^{-9}$, is the error induced by dipolar-coupling ($\kappa = 0.3$ kHz) between nearest neighbour NV centres; the second term, $p_{T_1} \sim (\Omega T_1)^{-1} \sim 10^{-4}$, is the depolarization error due to the finite NV spin relaxation time ($T_1 \sim 1$ ms); and the third term, $p_{T_2} \sim (\Omega T_2)^3 \sim 10^{-6}$, is the decoherence error due to the finite decoherence time ($T_2 = 100$ µs). The dominant contribution comes from the last term, $p_{\text{off}} = (\Omega_i/\Omega_R)^2 \sin^2(\Omega_R \tau/2)$, which is the crosstalk error induced by driving adjacent NV sites given finite NV ESR frequency detuning. Here $\Omega_R = \left(\Omega_i^2 + \Delta_{i,j}^2\right)^{1/2}$ is the generalized Rabi frequency and $i = $ A, …, D is the NV site label, with $\Omega_i$ being the on-resonant Rabi frequency at site $i$ and $\Delta_{i,j}$ being the ESR detuning between site $i$ and $j$. The maximum crosstalk error is observed to be 0.026(3) between NV site A and B with the following parameters: $\Omega_A$= 4.4(2) MHz, $\Omega_B$= 4.2(2) MHz, and $\Delta_{AB} = 27(1)$ MHz. See also Supplementary Fig. 11.

## Acknowledgements

This work was supported by the NSF, MURI QuISM, and DARPA QuASAR programs. We gratefully acknowledge the provision of diamond samples by Element 6 and helpful technical and theoretical discussions with Liang Jiang and Nathalie de Leon.

## Author Contributions

H.Z. and K.A. contributed equally to this work. R.L.W. conceived the idea of frequency encoding NV control and supervised the project. H.Z., K.A., and C.B. developed measurement protocols, hardware, and software, performed the measurements, and analysed the data. J.-C. J. carried out stimulated emission depletion (STED) imaging. All authors discussed the results and participated in writing the manuscript.

## Competing Financial Interests

The authors declare no competing financial interests.

**Figure legends**

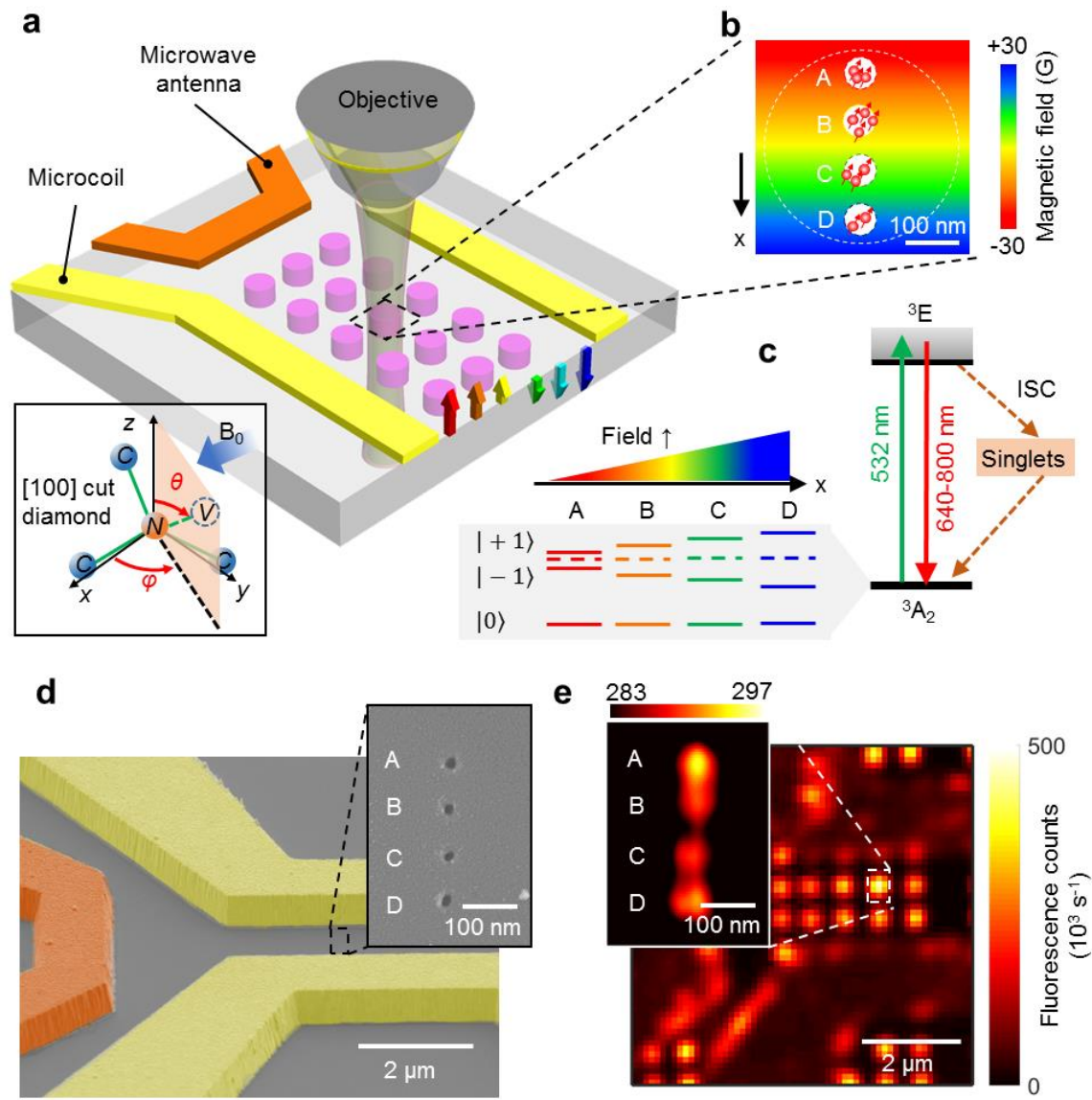

**Figure 1: Experimental set-up and characterization of NV centres and gradient microcoil.** **a**, Schematic of experimental apparatus. The diamond sample hosts negatively charged nitrogen vacancy (NV) colour centres implanted in a matrix of defined regions (pink circles) and integrated into a scanning confocal microscope. NV spin states in a given region are initialized and read out with a green (532 nm) laser, and manipulated with microwave fields produced by an antenna (orange bar). A uniform magnetic bias field of $B_0 = 128$ G, applied along one class of NV orientations, induces a Zeeman splitting between the $|\pm 1\rangle$ NV spin states, which are separated at zero magnetic field by 2.87 GHz from the $|0\rangle$ state. An additional gradient magnetic field (rainbow-colored arrows) of about 0.1 G nm$^{-1}$, created by electric current through a pair of gold wires (gradient microcoil), introduces a position-dependent Zeeman shift. **b**, Each region (pink circle in **(a)**) contains a 1 × 4 array of NV sites with 60 nm diameter and 100 nm centre-to-centre spacing, which can be exposed to the strong magnetic field gradient. Each site typically contains multiple (≈3±1) NVs of the selected orientation. **c**, NV energy-level diagram. Between the ground ($^3A_2$) and excited ($^3E$) electronic states there is a nonradiative intersystem crossing (ISC) channel. Due to the magnetic field gradient, the NV spin states $|\pm 1\rangle$ in each site acquire a position-dependent Zeeman splitting. By tuning the microwave frequency to a site-specific resonance, NV centres in any one site can be selectively addressed. **d**, Scanning electron microscope (SEM) image of gradient microcoil fabricated on a diamond substrate. The microcoil, represented by yellow pseudo-colour, is 1 µm thick and 2 µm wide. (Inset) SEM image of PMMA E-beam resist apertures used for ion implantation mask to create a 1× 4 array of NV sites. **e**, Scanning confocal microscope image of microcoil and matrix of regions hosting NV centres. (Inset) Stimulated emission depletion (STED) image of 1× 4 array of NV sites with 50 nm resolution. Colour table represents photon count rate in the unit of kilo-counts-per-second.

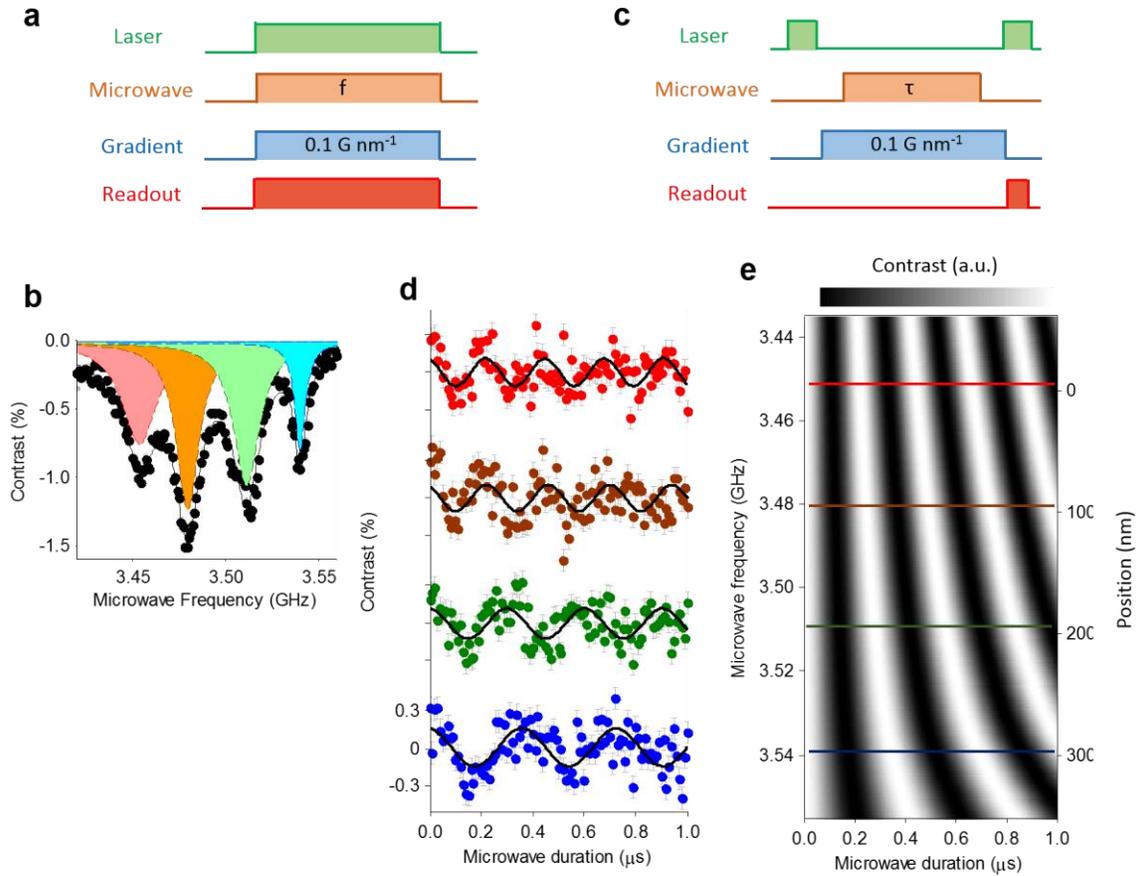

**Figure 2 | Site-selective NV ESR and Rabi oscillations. a**, Site-selective NV ESR sequence featuring continuous NV spin-state optical initialization and readout, a swept-frequency microwave field, and a fixed magnetic field gradient of 0.1 G nm$^{-1}$ (corresponding to $I = 250$ mA current). **b**, Measured NV ESR spectrum with four resonance peaks, corresponding to each NV site in a 1×4 array. Black solid line is a fit of the data (black dots) to a sum of four Lorentzian functions. The full-width-half-maxima of the peaks ≈8-15 MHz, and contrast ≈1.0-1.5 %. **c**, Site-selective NV Rabi driving sequence. Green laser pulses (5 μs duration) initialize and read out the NV spin states. A fixed magnetic field gradient (0.1 G nm$^{-1}$) differentiates the ESR frequencies at the four NV sites via Zeeman shifts. A microwave pulse, with frequency tuned to any one of the four ESR resonances, drives Rabi

oscillations at the corresponding NV site. **d**, Measured site-selective Rabi oscillation contrast as a function of microwave pulse duration $\tau_{MW}$. Black solid lines are fits of the data (coloured dots) to a sinusoid. From top to bottom, NV sites are driven at microwave frequencies of 3.452, 3.480, 3.511, and 3.540 GHz, with Rabi frequencies determined from fits to be 4.4(2), 4.2(2), 3.3(2), and 2.7(1) MHz, respectively. The Rabi frequency variation between sites is attributed to microwave field inhomogeneity caused by the boundary conditions of the gradient microcoil. **e**, Simulation with COMSOL of NV Rabi oscillation parameters provided by the gradient microcoil. The calculated Rabi frequency variation is consistent with the measurements in (**d**).

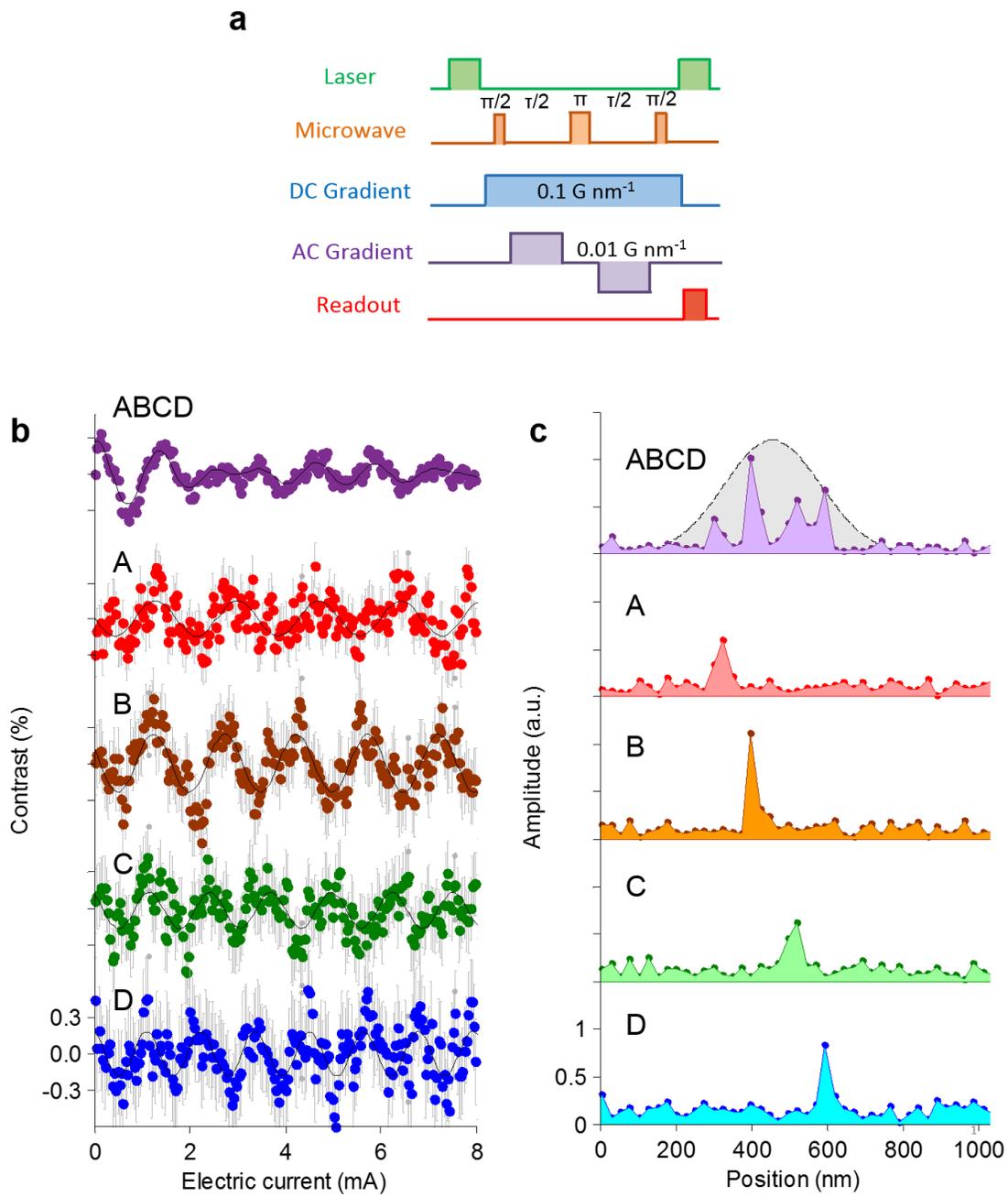

**Figure 3 | Site-selective NV Fourier imaging with nanoscale resolution. a**, Site-selective NV Fourier imaging sequence. NV spin states are initialized and read out with green laser pulses. A microwave Hahn echo ($\pi/2$- $\pi$- $\pi/2$) pulse sequence, synchronized with an alternating (AC) magnetic field gradient (amplitude ~0.01 G nm$^{-1}$), causes NV spins to

acquire position-dependent phase shifts, while also decoupling the spins from environmental noise. A fixed (DC) field gradient (0.1 G nm$^{-1}$) differentiates the ESR frequencies at the four NV sites via Zeeman frequency-encoding. **b**, One-dimensional k-space images of the four NV sites. The AC gradient strength and hence the wavenumber k are incrementally stepped by varying the current amplitude through the microcoil. The NV fluorescence is normalized to a reference measurement of the $|0\rangle$ state fluorescence, and a constant background level is subtracted. The top data set (labelled ABCD) is measured without the DC gradient, and thus all four NV sites are represented in the k-space image. The bottom four data sets (A, B, C & D) are measured with the DC gradient applied and the microwave frequency tuned to select only one each of the four NV sites. For all five k-space images, the solid black curves are fits of the data to a simple sinusoid model. **c,** One-dimensional real-space images of the four NV sites, obtained from the absolute value of the Fourier transforms of the k-space data. As with the k-space images, the top real-space image (ABCD) is with no DC gradient, and thus all four NV sites are represented; whereas the bottom four images (A, B, C & D) are with the DC gradient applied and the microwave frequency tuned to select only one each of the four NV sites. For comparison, the diffraction-limited, real-space point spread function of the confocal microscope is shown (grey shaded area, full-width at half-maximum of 300 nm). The location of each NV site determined from the site-selective images (A, B, C & D) agrees well with that obtained by Fourier imaging of all four sites without the DC gradient (ABCD).

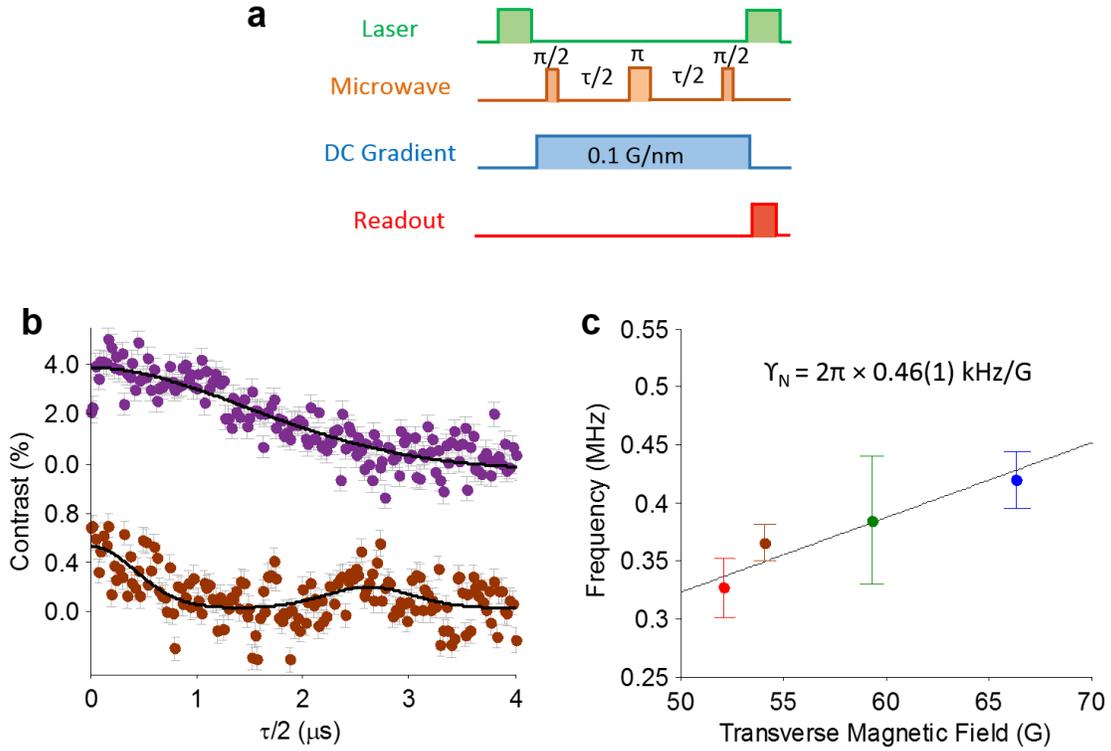

**Figure 4 | Site-selective NV NMR spectroscopy. a,** Site-selective AC magnetometry sequence, consisting of laser initialization & readout pulses, a microwave Hahn echo (π/2- π- π/2) pulse sequence with free precession time τ, and a DC field gradient (0.1 G nm$^{-1}$). The microwave frequency is tuned to one of the site-specific NV ESR resonances to be sensitive only to AC magnetic fields at the target spin site and with frequency ≈1/τ. **b,** Measured Hahn echo signal as a function of free precession time τ for no applied magnetic field gradient, which probes all four NV sites (upper panel); and an applied DC field gradient and the microwave frequency tuned to address only NV site C (lower panel). With no applied field gradient, NV spins in all sites experience only the uniform longitudinal bias field $B_0$; hence the NV Hahn echo signal provides a measure of the NV ensemble $T_2$ >2.5 µs. With an applied gradient, the NV Hahn echo signal is modulated by the Larmor precession of NV-constituent $^{15}$N nuclear spins in the transverse component of the gradient field $B_\perp$ at the selected NV site. In both cases, the data is well fit (black curves) by a model

that includes NV ESR and decoherence, $^{15}$N NMR, and the measurement protocol (see Supplementary Discussion 2). **c**, Measured $^{15}$N NMR frequency vs. transverse field $B_\perp$ at the four NV sites. $B_\perp$ values are determined from COMSOL simulations of the magnetic field gradient, constrained by measured NV ESR frequencies at each NV site. A linear fit to the data yields a $^{15}$N nuclear gyromagnetic ratio of $|\gamma_n/2\pi| = 0.46(4)$ kHz G$^{-1}$, which agrees with the accepted value [19].

# Supplementary Figures

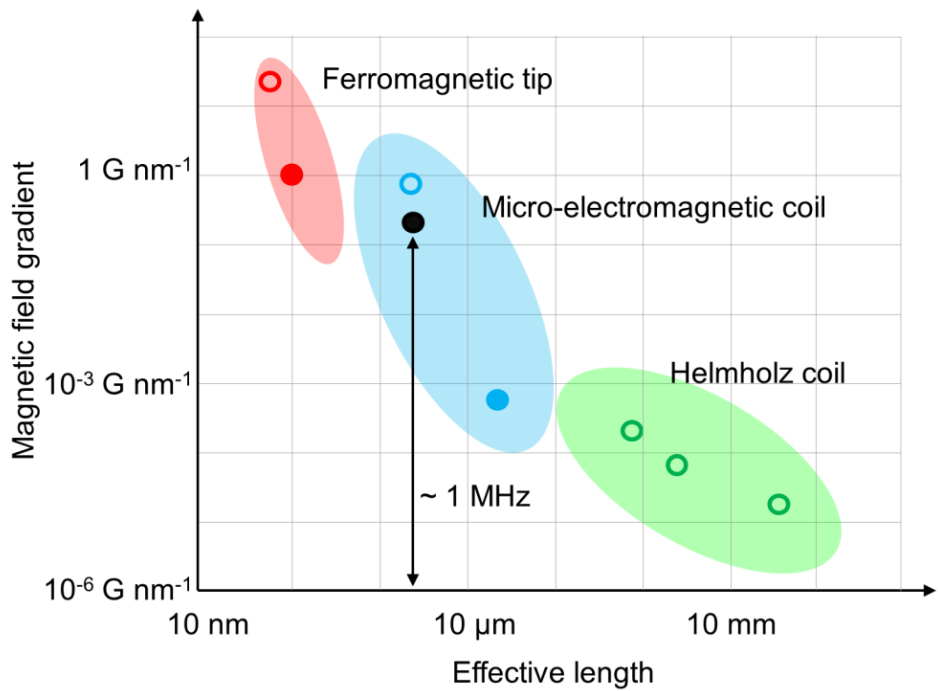

**Supplementary Figure 1 | Comparison of selected magnetic field gradient techniques.** Gradient strength as a function of effective length over which gradient magnitude applies. The open and filled circles indicate measurements at low temperature (< 77 K) and room-temperature, respectively. Ferromagnets (red) can produce the strongest gradients (>10 G nm$^{-1}$) at short length scales (<100 nm) [1-3]; however, switching the gradients requires slow mechanical motion of the magnet. Conventional Helmholtz and other 'volume' coils (green), as used in MRI [4-6], can create gradients ~10$^{-3}$ G nm$^{-1}$ (= 100 T m$^{-1}$) over 100 mm. Micro-electromagnetic coils (blue) offer intermediate performance [7, 8]. Our device (back dot) provides gradients >0.1 G nm$^{-1}$ that can be switched at ~1 MHz. The flexibility of microcoil design is useful for developing hybrid devices with microfluidic channels, Hall sensors, and microelectromechanical (MEMS) systems [9, 10]. These advantages make microcoils relevant for many applications, ranging from atomic manipulation [11, 12], to ferrofluid actuation [13], to manipulation of magnetotactic bacteria [14, 15] and DNA [16, 17].

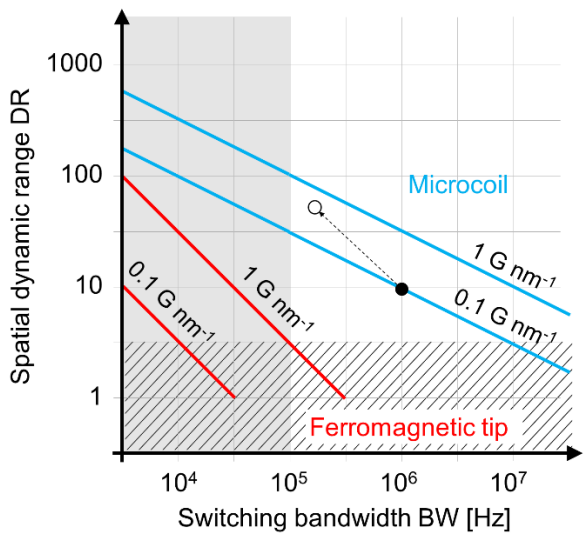

**Supplementary Figure 2 | Spatial dynamic range and gradient switching bandwidth for site-selective NV control experiments.** The spatial dynamic range is defined as DR = L/D, where L is the length scale across which the magnetic gradient is applied and hence NV sites can be selectively controlled, and D is the separation between individually addressable NV sites. DR < 2 (diagonal lines filled region) is not considered to be an array of distinct NV sites. The gradient switching bandwidth, BW, should be greater than the spin decoherence rate, which is ~100 kHz for typical dense NV samples (grey shaded region). The ferromagnetic tip and microcoil techniques are compared for two magnetic field gradient values, 0.1 G nm$^{-1}$ and 1 G nm$^{-1}$, corresponding to D ≈ 100 nm and 10 nm, respectively, for selective addressing of neighbouring NV sites with high fidelity. For a microcoil, DR and BW are limited by the microcoil size and electric current and related through DR ~ BW$^{-1/2}$ (Supplementary Discussion 4). The values demonstrated in this work are DR = 10, BW = 1 MHz (black filled circle). By increasing the current to 1.4 A, DR = 24, BW = 160 kHz can be achieved (black open circle). A ferromagnetic tip can be mechanically moved at a typical speed of 100 μm/s to switch to another site or turn off the gradient.

Thus, the switching rate is linearly proportional to the spatial dynamic-range (DR ~ $BW^{-1}$).

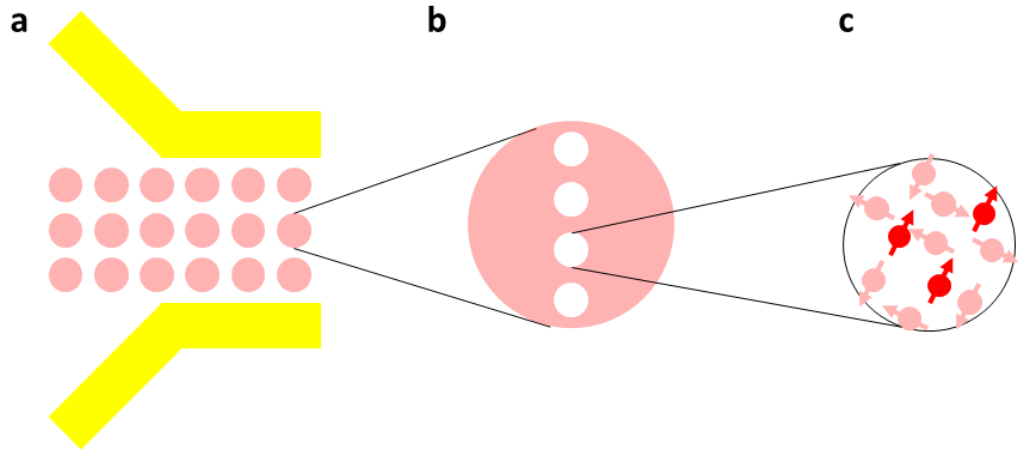

**Supplementary Figure 3 | Three-level hierarchical structure of 2D NV arrays.** The arrays of NV centres used in the present project can be described by a three-level hierarchical structure. **a**, The highest level is a 2D array of NV regions (~200 nm radius, 1 μm spacing), each of which can be individually optically addressed. **b**, The mid-scale level of a single NV region has four distinct NV sites (~30 nm radius, 100 nm spacing) aligned orthogonally to the microcoil wires, as shown in the zoomed-in view. Each NV site within a region can be selectively addressed via frequency encoding, as the NV ESR frequency becomes position dependent under the strong magnetic field gradient created by sending electric currents through the microcoil in an anti-Helmholtz configuration. **c**, At the smallest level, each site hosts multiple NV spins (typically 3±1 of a given orientation) with ~15 nm mean separation.

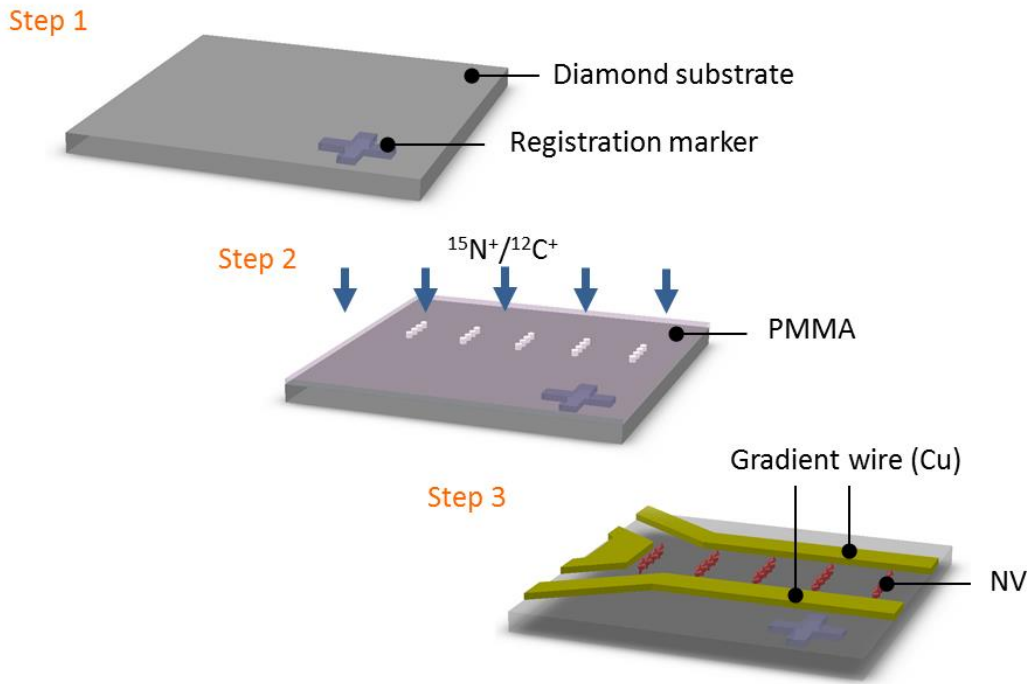

**Supplementary Figure 4 | Three steps for fabrication of frequency encoding device.**
Step 1: Registration markers (cross mark) for alignment are fabricated on a diamond substrate with e-beam lithography and reactive ion etching. All subsequent fabrication uses the spatial coordinates defined by these markers. Step 2: A polymethyl methacrylate (PMMA) ion implantation mask is used to spatially control NV centre formation in a three-level hierarchical structure. In this work $^{15}N^+$ ($10^{13}$ cm$^{-2}$, 14 keV) and $^{12}C^+$ ($10^{12}$ cm$^{-2}$, 20 keV) co-implantation is implemented to enhance conversion efficiency from $^{15}N^+$ ions to NV centres. Step 3: After NV centres are created via high-temperature annealing (1200 ºC, 4 hours), a Au microcoil is fabricated on the diamond and surrounding the 2D array of NV regions using e-beam lithography.

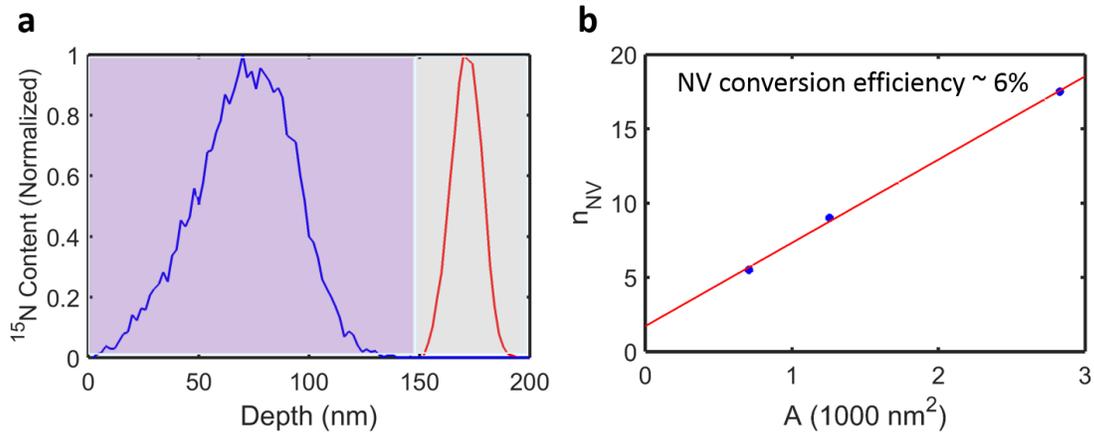

**Supplementary Figure 5 | Characterization of $^{15}$N implantation using PMMA mask. a,** Simulated spatial distribution of implanted $^{15}$N$^+$ ions as a function of depth using the Stopping and Range of Ions in Matter (SRIM) program. The blue and red curves show the spatial profile of the resulting $^{15}$N concentration formed by implanting $^{15}$N$^+$ ions through a PMMA mask (purple shaded area) and directly onto the diamond substrate (grey shaded area), respectively. When implanted through PMMA, more than 99.9% of $^{15}$N$^+$ ions are stopped within 150 nm from the PMMA surface and hence $^{15}$N$^+$ do not reach the diamond. The 150 nm thick PMMA (495 K C2) is formed with 3000 rpm spin coating. **b,** Number of NV centres ($n_{\text{NV}}$), determined from fluorescence count rate, is plotted as a function of the effective aperture area (A) with an ion implantation dosage of $10^{13}$ ions cm$^{-2}$. The estimated $^{15}$N$^+$ ion to NV centre conversion efficiency is 6%.

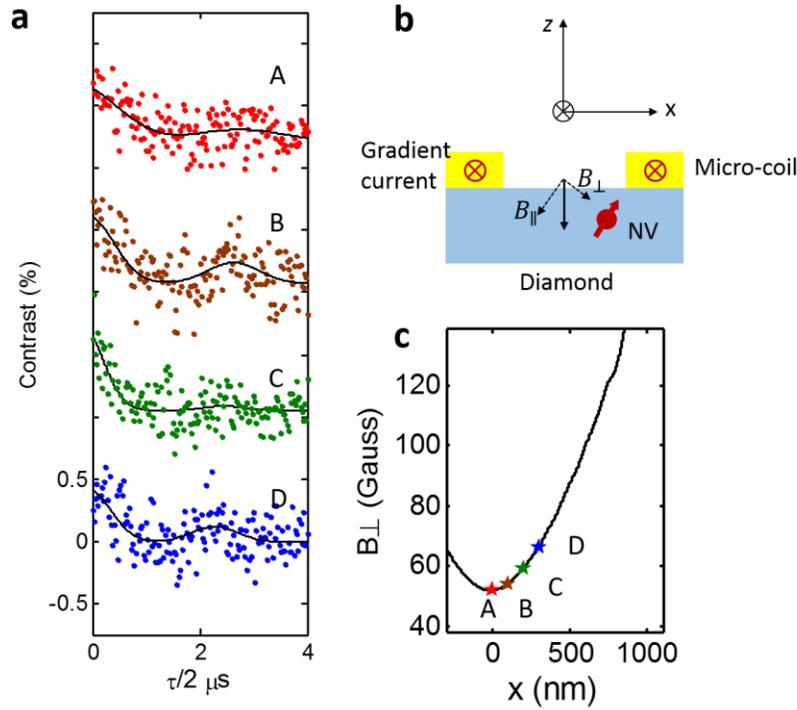

**Supplementary Figure 6 | Site-selective NV spin Hahn echo measurements with applied magnetic field gradient. a,** Measured NV Hahn echo signal as a function of half free precession time $\tau/2$ with applied DC field gradient (red, brown, green, and blue traces correspond to NV sites A, B, C, and D). For each trace, the microwave frequency is tuned to address only one NV site, yielding NV $T_2$ = 3.8±1.7, 4.6±1.1, 1.8±0.6, and 3.8±1.0 µs for sites A, B, C, and D, respectively. Modulation due to the $^{15}$N nuclear spin hyperfine interaction is observed, with different modulation frequencies across NV sites attributed to the varying transverse component of the applied DC field gradient ($B_\perp$). **b,** In the present experiment setup geometry, there is a finite angle between the NV crystallographic direction and the field gradient generated by the micro-coil, leading to $B_\perp$. **c,** COMSOL simulation of $B_\perp$ as a function of position (x). NV sites in site-selective Hahn echo measurements are labelled correspondingly.

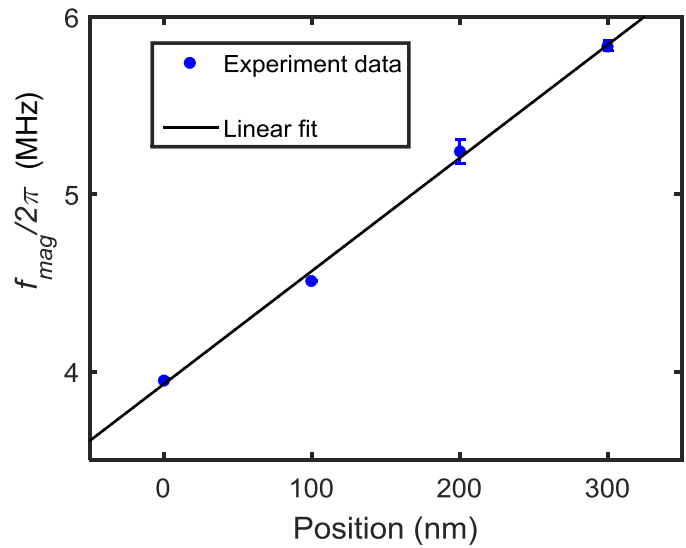

**Supplementary Figure 7 | Spatial frequency oscillations of site-selected 1D NV Fourier images.** From the 1D NV $k$-space images in Fig. 3b of the main text, the coherent single-spatial-frequency for each NV site is extracted by fitting to a sinusoidal function. Error bars are estimated from the 95% confidence intervals of fitting. The spatial frequencies are linearly proportional to the NV position, which indicates the uniformity of the magnetic field gradient as expected.

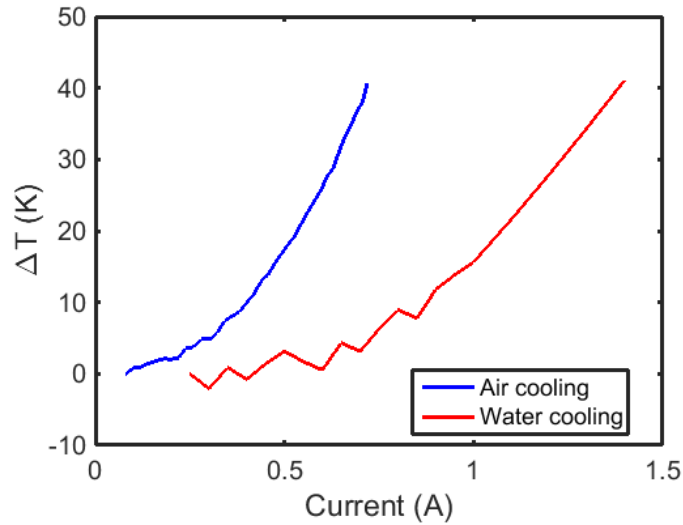

**Supplementary Figure 8 | Microcoil heat test.** To estimate the maximum current allowed in the field gradient device, the rise of microcoil temperature is measured as a function of electrical current by monitoring the change of resistance. When air cooling (shown as blue line) is used to dissipate heat from a heat sink attached to the NV diamond sample, the temperature increases by ~40 K with an electric current of 0.7 A, corresponding to a field gradient of ~0.3 G nm$^{-1}$. To further increase current, water is circulated on the heatsink. With the water cooling (red), the microcoil tolerates approximately twice the current (1.4 A), corresponding to a field gradient of ~0.6 G nm$^{-1}$.

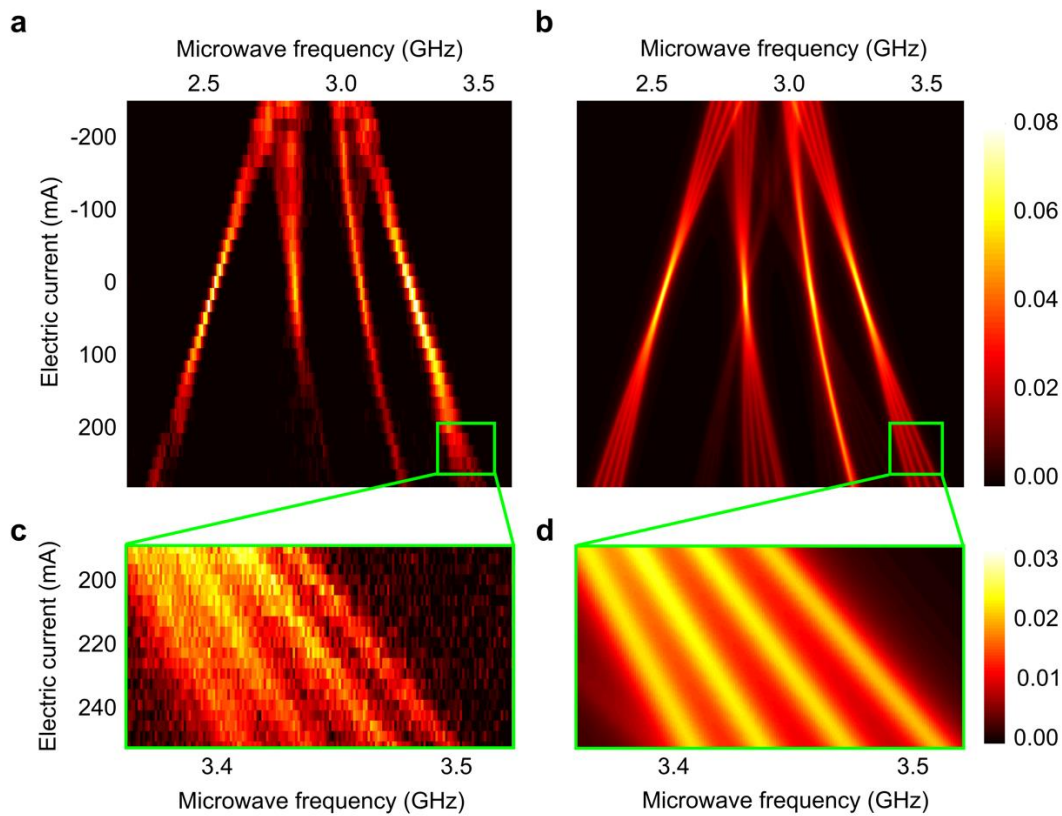

**Supplementary Figure 9 | Effect of magnetic field gradient on NV ESR spectrum**. **a**, Two-dimensional NV ESR spectrum measured by sweeping both the electric current through the microcoil (vertical axis) and the microwave driving frequency (horizontal axis). A permanent magnetic field of $B_0 = 128$ G along one class of NV orientations produces four bands of resonance peaks. The outermost and innermost two bands correspond to the transitions of the on-axis and three degenerate off-axis NV centres, respectively. As the electric current is increased, the resonance bands move with respect to each other. In particular, the inner two bands start to split into six. At larger current, only four bands out of eight have large contrast because the laser is linearly polarized and hence couples mostly to two NV axes. **b**, Simulated NV ESR spectrum. The magnetic field pattern is calculated at a fixed current of $I = 250$ mA using COMSOL. The field is assumed to be linearly proportional to the current. With the simulated field as an input, ESR resonance frequencies for all four NV orientations and four NV sites are determined by diagonalising the NV

Hamiltonian. **c**, Finer spectral resolution scan of the NV ESR spectrum with the sweep range indicated by the blue box in (**a**). The stronger the magnetic field gradient, the larger the splitting of ESR lines between adjacent NV sites (100 nm separation). **d**, Simulation corresponding to (**c**) and zoomed-in view of blue box in (**b**).

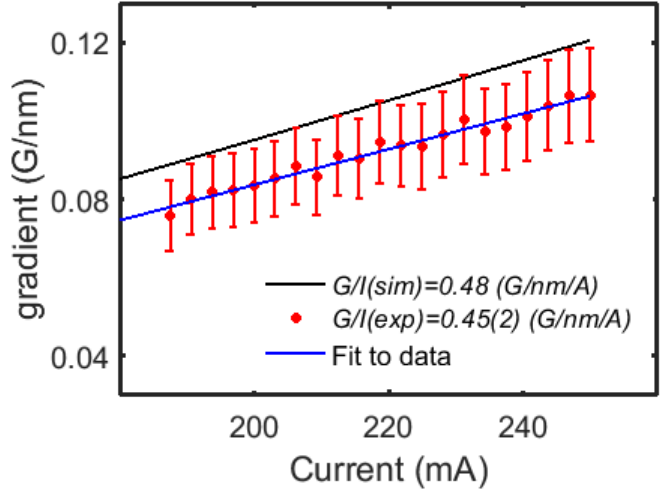

**Supplementary Figure 10 | Magnetic field gradient as a function of microcoil current.** NV ESR measurements are used to determine the magnetic field gradient as a function of electrical current (red dots), with a linear fit to the data (blue line) yielding the field gradient per unit current as $dB/dx/I = 0.45(2)$ G nm$^{-1}$ A$^{-1}$. A numerical simulation of the field gradient produced by the microcoil (black line) gives $dB/dx/I = 0.48$ G nm$^{-1}$ A$^{-1}$. The disagreement between measurement and simulation could be attributed to the uncertainty of NV orientation with respect to the microcoil. Error bars on the data points are inferred from two factors: (i) 67% confidence interval of fitting each NV ESR spectrum to a sum of four Lorentzian lines; and (ii) uncertainty of NV centre position within each site in the array. In particular, the NV centres are assumed to be created randomly within the 60 nm diameter aperture of each array site, giving the standard deviation of NV location from the centre as $\sigma = 17.3$ nm.

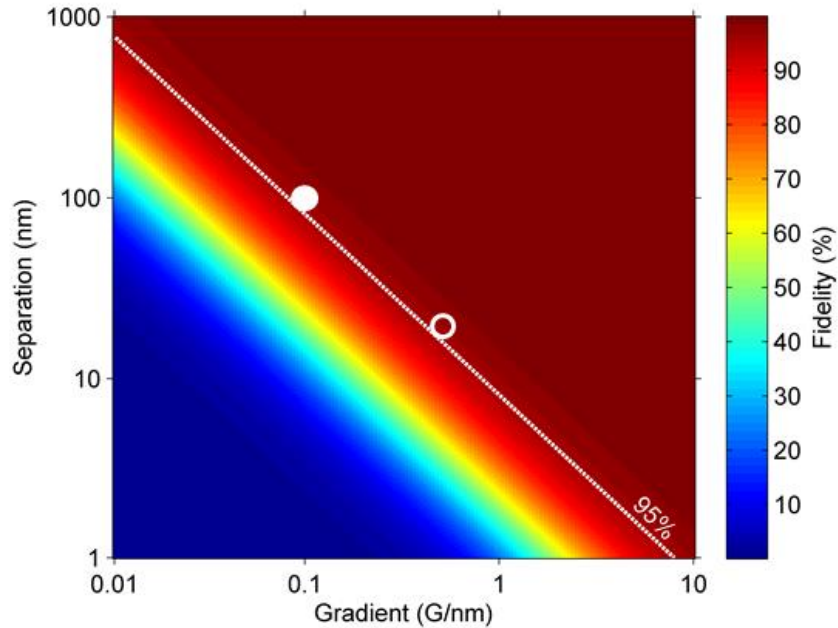

**Supplementary Figure 11 | Projected fidelity of site-selective NV Rabi driving.** The fidelity of site-selective NV spin driving in the presence of a magnetic field gradient is calculated from $\Delta^2/(\Delta^2 + \Omega^2)$, where $\Omega = 5$ MHz is the typical Rabi driving frequency, and $\Delta = \gamma D * dB/dx$ is the frequency detuning of the Rabi drive field between NV sites calculated from the gradient *dB/dx* and NV site separation D. The values used in this work are *dB/dx* = 0.1 G nm$^{-1}$ and D = 100 nm (white filled circle), indicating a fidelity >95%. By water cooling, we expect to be able to drive NV sites separated by D = 20 nm using a stronger gradient of *dB/dx* = 0.6 G nm$^{-1}$ (white open circle).

# Supplementary Discussion

## 1. Microwave inhomogeneity

In Fig. 2d, a variation in Rabi frequency is observed across the four NV sites, which can be explained by the non-uniform microwave field distribution created by the antenna. The antenna + NV array system is in the microwave near-field; hence the spatial variation of the microwave field is determined by the dimensions of surrounding conductors rather than the wavelength of the radiation. Since the electric field is zero ($\mathbf{E} = 0$) in a conductor, no time-varying magnetic field exists in the microcoil according to the Maxwell-Faraday equation $\mathbf{\nabla} \times \mathbf{E} = -\partial \mathbf{B}/\partial t$. Across the boundary between the microcoil and the surrounding medium, the normal component of the magnetic field $B_y$ should be continuous. As a result, only a tangential component of the microwave field $B_x$ exists near the microcoil boundary. This boundary effect alters the magnitude as well as the direction of the microwave field in the near field, which leads to a significant change in the NV Rabi frequency across the dimensions of the four-site array. Note that such microwave field gradients have also been used for selective addressing and motional-internal state coupling of atomic systems [19, 20].

## 2. Transverse magnetic field induced by gradient microcoil

The gradient field produced by the microcoil is nominally aligned with the uniform static field $B_0$ in order to perform NV frequency encoding and Fourier imaging. There is also a small transverse component to the gradient field $B_\perp(x)$, orthogonal to the direction of $B_0$, which is exploited for site-selective NMR spectroscopy of $^{15}$N nuclear spins (Fig. 4 and Supplementary Fig. 6). Thus, for the experimental protocol shown in Fig. 4a, the envelope of the NV Hahn echo signal in the presence of a DC magnetic field gradient is modulated as $S(\tau) \propto \sin^2(\omega_A t/2) \sin^2(\omega_B t/2)$, where $\omega_A \approx 2\pi \times 3$ MHz is the NV-$^{15}$N hyperfine coupling strength and $\omega_B \approx 14\gamma_n B_\perp$ is the $^{15}$N nuclear spin Larmor frequency. Here $\gamma_n = 2\pi \times 0.43$ kHz G$^{-1}$ is the gyromagnetic ratio of the bare $^{15}$N nuclear spin, and the factor of 14 is an effective enhancement of this gyromagnetic ratio due to virtual transitions arising from higher order NV-$^{15}$N hyperfine interactions [18]. From the observed modulation frequency $\omega_B = 2\pi \times 0.36(2)$ MHz, the transverse field at NV site C is determined to be $B_\perp = 59(4)$ G, which agrees well with a COMSOL simulation of the gradient field produced by the microcoil (Supplementary Fig. 6c). Similar agreement between measurement and simulation is found for the other NV sites in the array.

### 3. Extension of selective addressing to multiple NV sites and spatial dimensions

In future work, straightforward methods can be used to extend selective control to two or more NV sites as well as a second spatial dimension. For example, multiple NV sites could be addressed simultaneously, in the presence of a frequency-encoding DC magnetic field gradient, by applying a microwave signal with spectral components at the NV ESR frequencies of all target sites. To selectively address a two-dimensional (2D) array of NV sites (e.g., a square lattice), one cannot simply employ simultaneous frequency-encoding gradients in both dimensions because a degeneracy of the resonance frequency occurs. Instead, one can maintain the spin polarization of a target row of NV sites, while dephasing the other rows; and then performing 1D frequency encoding along the target row to select the desired site. This procedure can be realized, for instance, by (i) putting all NV spins in the 2D array onto the equator of the Bloch sphere with no DC gradient applied and a uniformly resonant microwave $\pi/2$ pulse; (ii) bringing only the target $x = x_0$ row back to the initial polarized state with a frequency-encoded $3\pi/2$ microwave pulse in the presence of a DC field gradient; (iii) waiting for $T_2^*$ dephasing of the other (non-target) NV rows; and then (iv) performing 1D frequency encoding in the y direction to select the target NV site $(x_0, y_0)$.

## 4. Microcoil performance

*Maximum gradient strength and switching bandwidth*

Though the frequency encoding demonstrations reported in the present work are performed with a maximum current of 250 mA, the gradient microcoil can be operated at a much higher current (and hence larger gradient strength) before it breaks down due to ohmic heating. The maximum current that the microcoil can support is determined by measuring the device temperature rise as a function of electric current (Supplementary Fig. 8). With water cooling, the temperature rises by nearly 40 K with $I = 1.4$ A, with a corresponding gradient of 0.6 G nm$^{-1}$. The switching bandwidth of the gradient pulse is determined by measuring the rising NV Rabi signal contrast as a function of delay between the initiation of the gradient pulse and $\pi/2$ pulse. Using impulse response function theory, the 3 dB switching bandwidth is $f_{3dB} = (\ln(0.9)-\ln(0.1))/(2\pi Tr) \sim 0.9$ MHz, where $Tr = 400$ ns is the measured 10%-90% rise time in the NV Rabi signal contrast.

*Relation between spatial dynamic range and switching bandwidth*

To characterize the maximum number of individually addressable NV sites, the spatial dynamic range of site-selective control, DR = L/D is introduced. Here L is the length scale across which the strong gradient is applied, and D is the separation between individually addressable NV sites. For a microcoil, L is determined by the separation between the two wires and D is inversely proportional to the gradient strength. A magnetic field gradient of $dB/dx = 0.1$ G nm$^{-1}$ and 1 G nm$^{-1}$ corresponds to D ≈ 100 nm and 10 nm, respectively, assuming a Zeeman shift of about 30 MHz is needed for selective addressing of neighbouring NV sites with high fidelity. Since the gradient strength decreases quadratically as the coil separation increases ($dB/dx \sim I/L^2$) for a fixed value of gradient strength or D, the electric current $I$ should be increased by $I \sim L^2 \sim DR^2$. However the maximum achievable current $I$ is also limited by the switching bandwidth, i.e., $I \sim BW^{-1}$. BW can be affected by at least 3 factors: **a**, self-inductance of the micro-coil; **b**, peripheral circuits, such as in the printed circuit board (PCB) that interfaces the current source and micro-coil; and **c**, finite speed of the pulsed current source, which is a current amplifier in our case. COMSOL simulation indicates that the micro-coil has a very small self-inductance ~ 1 nH, and hence the corresponding bandwidth limit is on the order of GHz. Finite bandwith due to peripheral circuits can be, in principle, eliminated by integrating the current source close to the micro-coil. So the current amplifier is expected to be the primary switching bandwidth limitation. Furthermore, assuming that the output from the current amplifier is slew rate (SR) limited (which is usually the case when high current output or large voltage is needed), BW can be expressed as: $BW = SR/(2\pi V_p)$, where $V_p$ stands for peak output voltage [21]. Since current is proportional to voltage, then BW and current are related as $I \sim BW^{-1}$. Thus we conclude that the spatial dynamic range and switching bandwidth are related via $DR \sim BW^{-1/2}$ (Supplementary Fig. 2).

### 5. Outlook for fabrication of high NV density arrays

High-precision nitrogen ion implantation is a key technical challenge for preparing either single NV centre arrays or strongly-coupled dense NV spin baths. Controlling the separation and number of NV centres with high precision remains an active area of research. It is thus of importance to evaluate the minimum limits of NV site area (A), NV-NV separation within a single site (d), and spacing between sites (D), at the present state of technology. The NV site area is constrained by the mask diameter and lateral implantation straggle. Using a polymethyl methacrylate (PMMA) mask, apertures with diameter ≈ 30-60 nm can be created. Further reduction to 10 nm by reducing the mask thickness or even to ~1 nm by combining with atomic layer deposition of alumina may be possible; however, the lateral implantation straggle (7.0 nm at 14 keV implantation energy) will eventually limit the NV site area to $A_{min}$ ≈ $(10\ nm)^2$. Nevertheless, this is small enough for quantum information processing architectures based on spin chains with error correction methods [22, 23]. The NV-NV separation within a single NV site is determined by the NV density. With an $^{15}N^+$ implantation dosage of $10^{13}$ $cm^{-2}$ and N-to-NV conversion efficiency of 6 %, the diamond sample used in the present work has d = 15 nm. By increasing the implantation dosage and irradiating with neutrons or electrons, a minimum separation of $d_{min}$ ≈ 4 nm (corresponding to an NV-NV dipolar interaction strength ≈ 0.6 MHz) may be achievable, although the sample at this density might suffer from low fluorescence contrast due to $NV^-$/$NV^0$ charge state conversion and a shorter $T_2^*$ due to remaining paramagnetic impurities. The spacing between NV sites, on the other hand, is limited by the NV site area, and thus cannot be smaller than $D_{min}$ ≈10 nm. If two NV centres are randomly picked from each of two NV sites with area A and spacing D, the mean and standard deviation of separation between them are given by ~D and ~$A^{1/2}$, respectively.

## Supplementary References